\documentclass[onecolumn]{rmaa}

\title{On the Distance to the Bright Nonthermal Radio Sources in the
Direction of an Extraordinarily Massive Cluster of Red Giants}

\author{Alfonso Trejo
and
  Luis F. Rodr\'\i guez
  \affil{Centro de Radioastronom\'{\i}a y Astrof\'{\i}sica, UNAM, Morelia} 
}

\fulladdresses{
\item Alfonso Trejo and 
Luis F. Rodr\'\i guez: Centro de Radiostronom\'{\i}a
y Astrof\'\i sica, 
Universidad Nacional Aut\'onoma de M\'exico, A. P. 3-72, 
(Xangari), 58089 Morelia, Michoac\'an, M\'exico
(a.trejo; l.rodriguez@astrosmo.unam.mx).
} 

\shortauthor{Trejo \& Rodriguez}
\shorttitle{Radio Sources in the Direction of an Extraordinarily Massive Cluster}

\SetVolume{42} \SetFirstPage{000} \SetYear{2006}
\ReceivedDate{2006 March 22} 
\AcceptedDate{2006 April 06} 

\resumen{Un c\'umulo extraordinariamente masivo de
supergigantes rojas ha sido recientemente reportado en la direcci\'on 
de las coordenadas gal\'acticas 
$l = 25\rlap.^\circ3$; $b = -0\rlap.^\circ2$.
Este c\'umulo est\'a asociado con una fuente de rayos X,
con una fuente de muy alta energ\'\i a de
rayos $\gamma$, y con tres fuentes de radio brillantes y
no t\'ermicas. 
La pro\-ba\-bilidad \sl a priori \rm de que estas asociaciones
sean s\'olo coincidencias en la l\'\i nea de visi\'on es
muy peque\~na. Sin embargo, hemos analizado datos del
archivo del VLA y encontramos del espectro de absorci\'on de HI 
de las tres fuentes de radio que son extragal\'acticas y
por lo tanto no asociadas directamente con el c\'umulo gal\'actico.
}
 
\abstract{An extraordinarily massive cluster of red supergiants has
been recently reported in the direction of the galactic coordinates
$l = 25\rlap.^\circ3$; $b = -0\rlap.^\circ2$. 
This cluster is associated with an X-ray source, a very high-energy $\gamma$-ray 
source, and three bright non-thermal radio sources. The \sl a priori \rm
probability of these associations being only line-of-sight
coincidences is very small. However, we have analyzed VLA archive data 
taken toward these three radio sources and find from
their HI absorption spectra that they are extragalactic and thus not
directly associated with the galactic cluster.
}
      

\begin{document}

\maketitle

\section{Introduction}

Recently, Figer et al. (2006) reported the discovery of an extraordinarily 
massive young cluster of stars in the Galaxy, having an
inferred total initial cluster mass of 20,000 to 40,000 $M_\odot$,
comparable to the most massive young clusters
in the Galaxy.
Using {\it IRMOS}, {\it 2MASS}, and {\it Spitzer} observations, they concluded 
that there are 14 
red supergiants in the cluster, compared with five in NGC~7419
(Caron et al. 2003), 
previously thought to be the richest Galactic cluster of such stars.

\begin{figure*}[htb]
\centering
  \includegraphics[width=1.0\columnwidth]{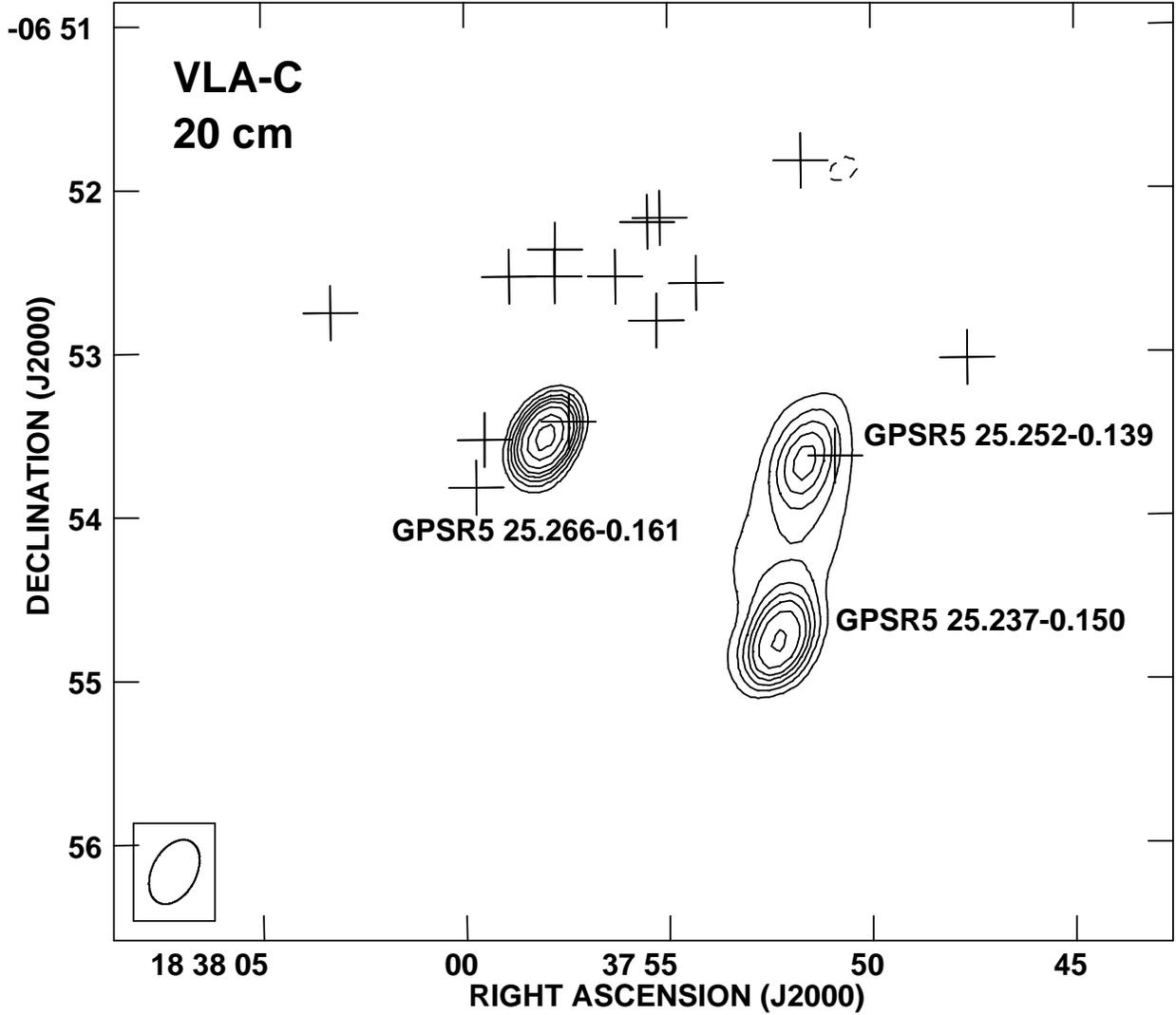}
  \caption{VLA contour image of the continuum emission at 20 cm
of the region of the cluster of red supergiants. Contours are
-4, 4, 6, 8, 10, 12, 15, 20, and 25 times 13 mJy beam$^{-1}$, the
rms noise of the image. The image was made from line-free channels of the
data and it is corrected for the primary beam response. The names of the
three bright non-thermal sources are indicated in
the figure. The positions of the fifteen
red supergiants reported by Figer et al. (2006) are marked with crosses.
The half-power contour of the synthesized beam 
($25\rlap.{''}3 \times 16\rlap.{''}1$ ; PA = $-28^\circ$) is shown
in the bottom left corner of the map.}
  \label{fig:cluster}
\end{figure*}

\begin{table*}[htbp]
\small
  \caption{Bright Radio Sources Near the Massive Cluster of Red Supergiants}
  \begin{center}
    \begin{tabular}{lcccccc}\hline\hline
&\multicolumn{2}{c}{Position} & 20-cm Flux  &HI spectrum box 
&Proposed \\
\cline{2-3} 
Source &  $\alpha$(J2000) & $\delta$(J2000) & Density (Jy) & 
($\Delta \alpha \times \Delta \delta$) & 
Identification \\ 
\hline
GPSR5~25.252-0.139& 18 37 51.7 & -06 53 40 & 0.38$\pm$0.04 & $24'' \times 40''$ 
& Background Extragalactic \\
GPSR5~25.237-0.150& 18 37 52.3 & -06 54 46 & 0.44$\pm$0.04 &  $16'' \times 24''$ 
& Background Extragalactic \\
GPSR5~25.266-0.161& 18 37 58.0 & -06 53 31 & 0.37$\pm$0.04 & $16'' \times 24''$ 
& Background Extragalactic  \\
G25.38-0.18       & 18 38 15.1 & -06 48 01 & 3.92$\pm$0.13 & $24'' \times 40''$  
&  Galactic H~II Region \\

\hline\hline
    \label{tab:1}
    \end{tabular}
  \end{center}
\end{table*}
Furthermore, they find that the cluster is associated with an x-ray source 
(detected by {\it ASCA} and {\it Einstein}), a recently discovered very high energy $\gamma$-ray 
source (detected by {\it INTEGRAL} and {\it HESS}), and several non-thermal radio sources, 
suggesting that these objects are likely related to recent supernovae
in the cluster. The non-thermal sources are quite bright and an HI absorption spectrum toward
them could be used to estimate their distance. Here we present the analysis
of VLA archive data that allows such a distance determination. 

\section{Observations and Discussion}

\begin{figure*}
\centering
\vskip-0.7cm
\includegraphics[scale=0.4, angle=0]{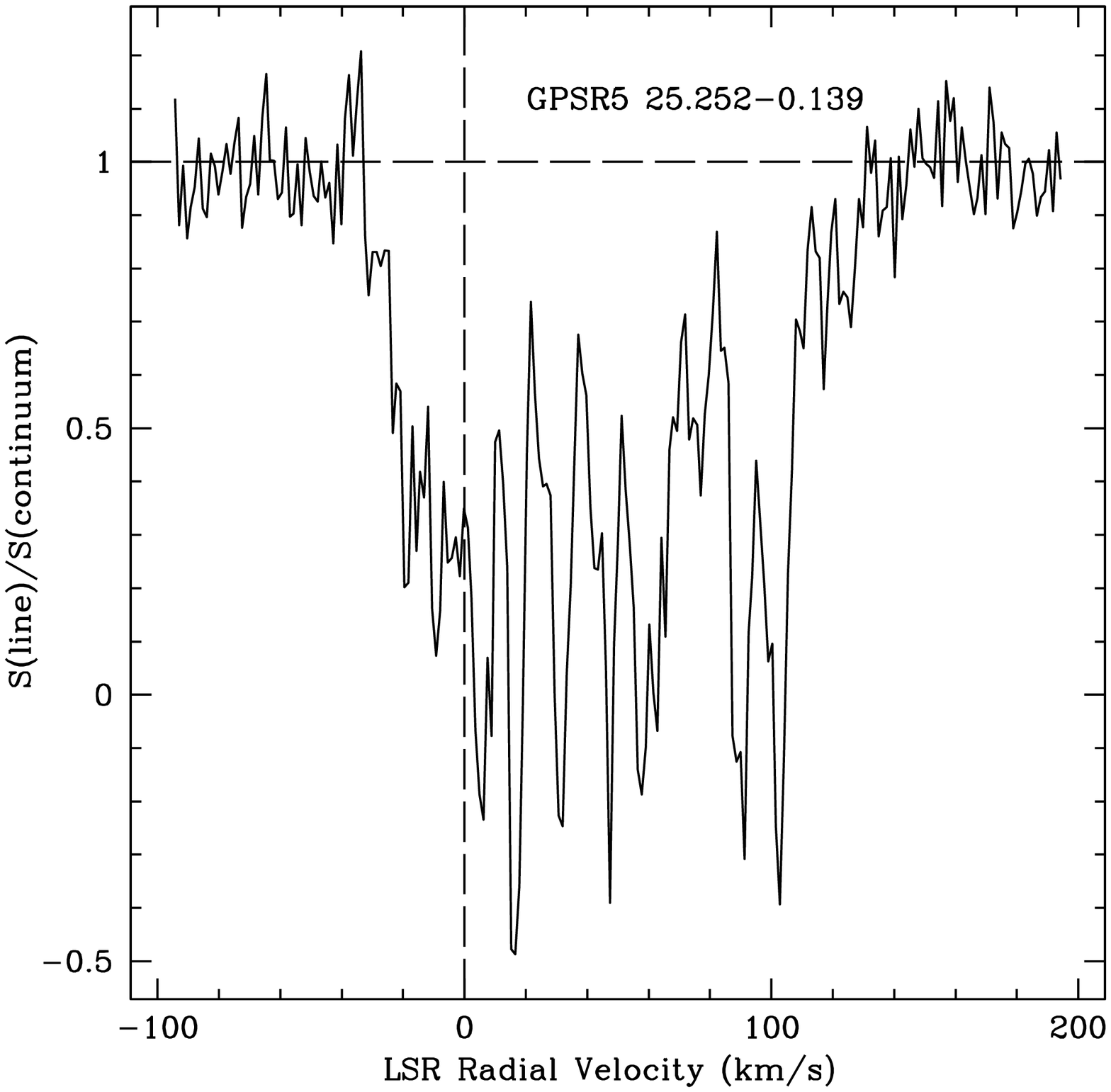}%
\hspace*{\columnsep}%
\includegraphics[scale=0.4, angle=0]{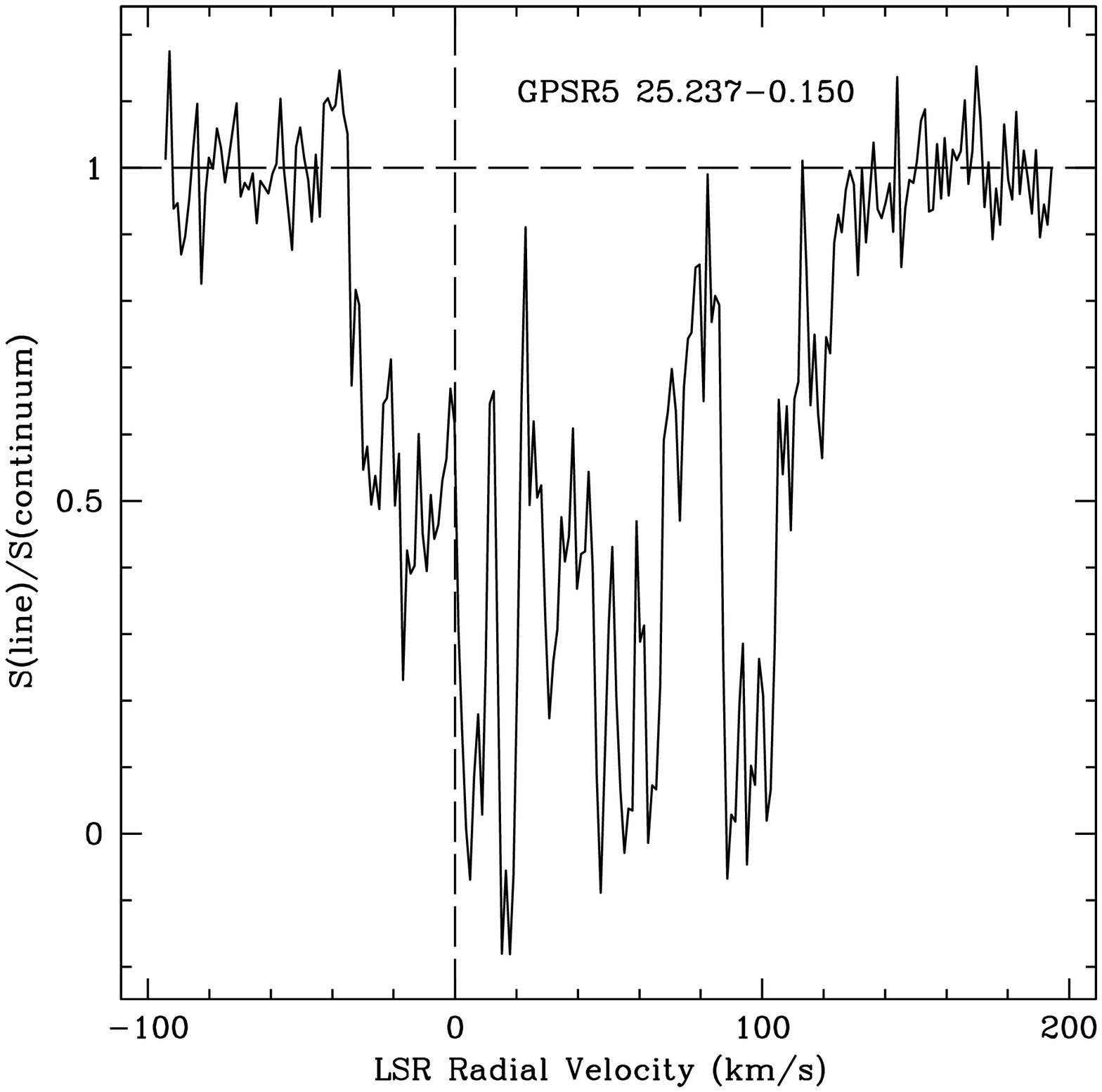}
\includegraphics[scale=0.4, angle=0]{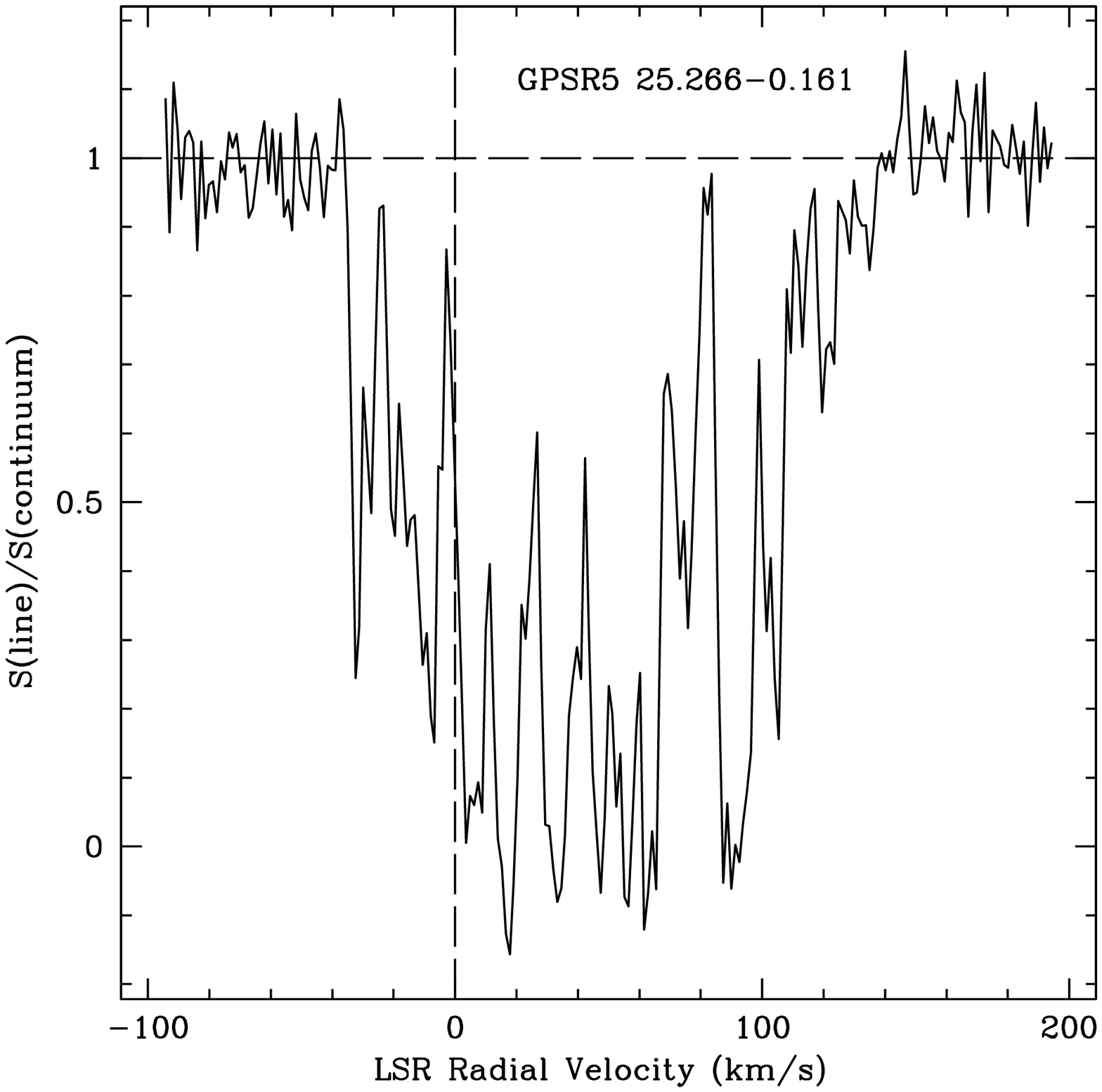}%
\hspace*{\columnsep}%
\includegraphics[scale=0.4, angle=0]{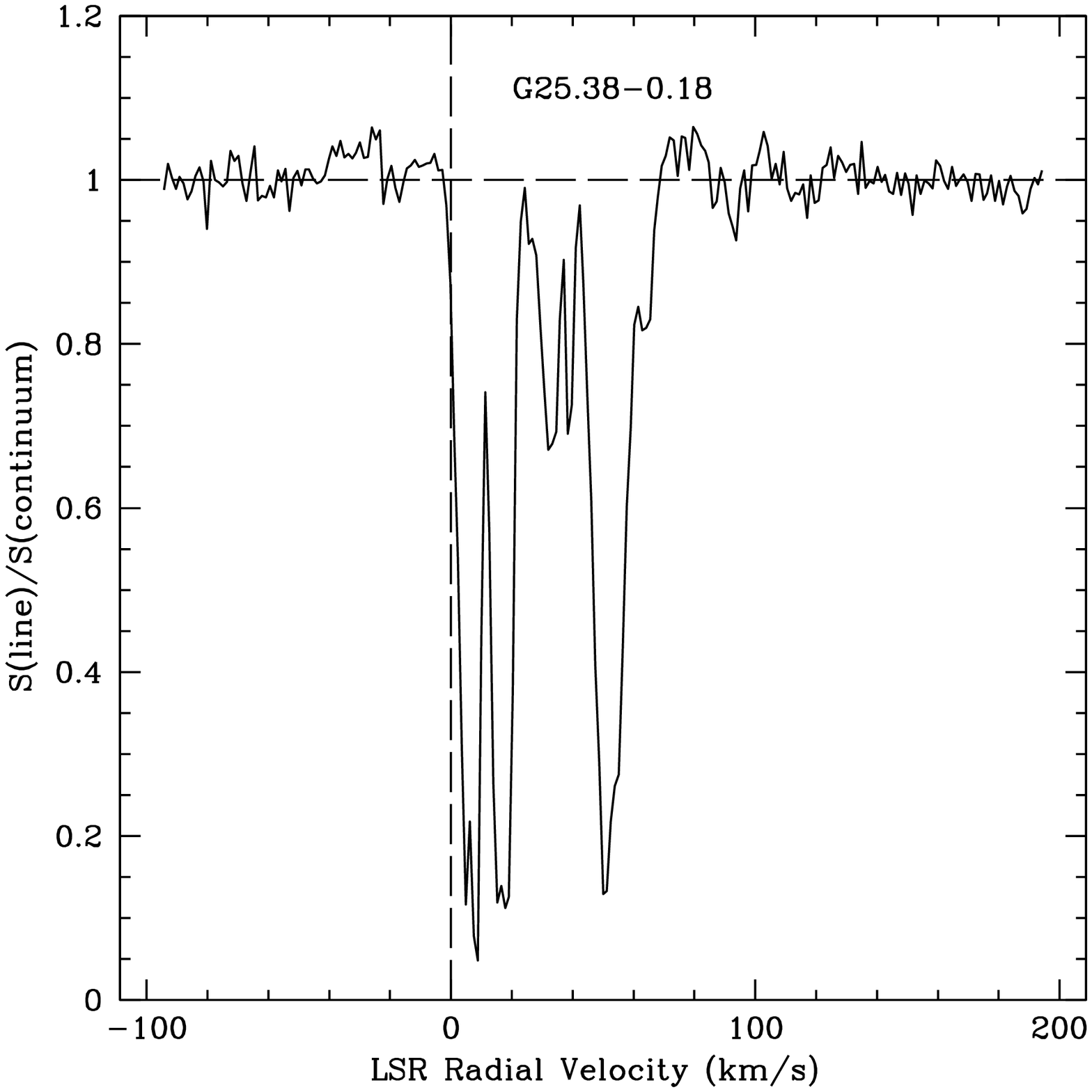}
 \caption{21 cm H I absorption spectra for the four sources listed in
Table 1, plotted as a function of LSR radial velocity. The measured spectra was divided 
by the continuum level and represents exp(-$\tau$).
The horizontal dashed line is drawn at $S(line)/S(continuum)$ = 1
and the vertical dashed line is drawn at an LSR radial velocity
of 0 km s$^{-1}$.
Note that in the three non-thermal sources (GPSR5~25.252-0.139,
GPSR5~25.237-0.150, and GPSR5~25.266-0.161) there is significant absorption 
from gas beyond the solar circle (negative velocities).
In contrast, the H~II region G25.38-0.18 does not show
absorption at negative velocities, indicating that the source is located
inside the solar circle.
}
  \label{fig2}
\end{figure*}

The HI and 20 cm continuum observations were taken
from the archive of the Very Large Array (VLA)
of the NRAO\footnote{The National Radio
Astronomy Observatory is operated by Associated Universities
Inc. under cooperative agreement with the National Science Foundation}.
The observations
were taken on 2000 April 6. The VLA 
was then in the C configuration, providing an angular resolution of about 
$20{''}$ for images made with natural weighting.
The absolute amplitude calibrator was 1331+305 
(with an adopted flux density of 14.7 Jy at 1.4 GHz). 
and the phase calibrator was 1743$-$038 , with a bootstrapped 
flux density of 2.18$\pm$0.01 Jy. 
The data were edited and calibrated using 
the software package Astronomical Image Processing System (AIPS) of NRAO. 

\subsection{Continuum}

A natural-weight continuum image, made using the line-free channels of the data, is shown
in Figure 1. The three continuum sources, as well as their close association
with the supergiant stars, are evident in this image. The spectral indices
of these sources indicate a non-thermal nature (Becker et al. 1994).
The positions and flux densities
of the sources are given in Table 1. In addition to the three non-thermal
sources, we have also included in this Table the well known H~II region
G25.38-0.18 (e. g. Wink, Wilson, \& Bieging 1983;
Churchwell, Walmsley, \& Cesaroni 1990) that 
appears projected about $8'$ to the NE of the cluster.
This H~II region is the brightest component of the W42 complex.
We include this source in the discussion to have an HI absorption spectrum of
a confirmed galactic source for comparison.

The \sl a priori \rm probability of finding a background source with a 20-cm
flux density of $\sim$0.4 Jy (the average flux density
of the three non-thermal sources) in a region of about $4' \times 4'$ is very small.
Following Windhorst et al. (1993), we estimate this probability to be only $\sim$0.0005. 
This suggests a real association between the cluster and the radio sources.

\subsection{HI Line}

However, these three non-thermal radio sources are included among the
54 bright, compact radio sources used by Kolpak et al. (2002) in their
study of the radial distribution of cold hydrogen in the Galaxy. 
Although these authors do not show HI absorption spectra of these sources, they note
that all sources considered in their analysis are identified as 
extragalactic continuum sources by the presence of significant absorption at 
local standard of rest (LSR) velocities less than $-$10 km s$^{-1}$. 
This negative velocity absorption is caused by relatively remote material
in the Galaxy, located outside the solar circle. 

To investigate in
detail the HI absorption spectra, we made line cubes of the region and produced
spectra toward the sources in an angular box containing most of the
bright central emission of the sources. The dimensions of these boxes
are given in Table 1. The spectra has a velocity resolution of 
2.6 km s$^{-1}$ and they are shown in
Figure 2.

As can be seen in Figure 2, the three non-thermal sources
(GPSR5~25.252-0.139,
GPSR5~25.237-0.150, and GPSR5~25.266-0.161) show HI absorption in
an LSR velocity range of $\sim$$-$30 to +120 km s$^{-1}$, while the
H~II region G25.38-0.18 shows HI absorption in the more limited range
of $\sim$0 to +60 km s$^{-1}$.
To interpret the implications of these absorption spectra,
we use the galactic rotation model of Brand \& Blitz (1993)
and assume that the HI disk of the Galaxy has an outer radius of 13.4 kpc (Goodwin, 
Gribbin, \& Hendry 1998).
In Figure 3 we show the expected LSR velocity as a function of distance
to the Sun in the direction of the cluster of red supergiants.

\begin{figure}[htb]
\centering
  \includegraphics[width=1.0\columnwidth]{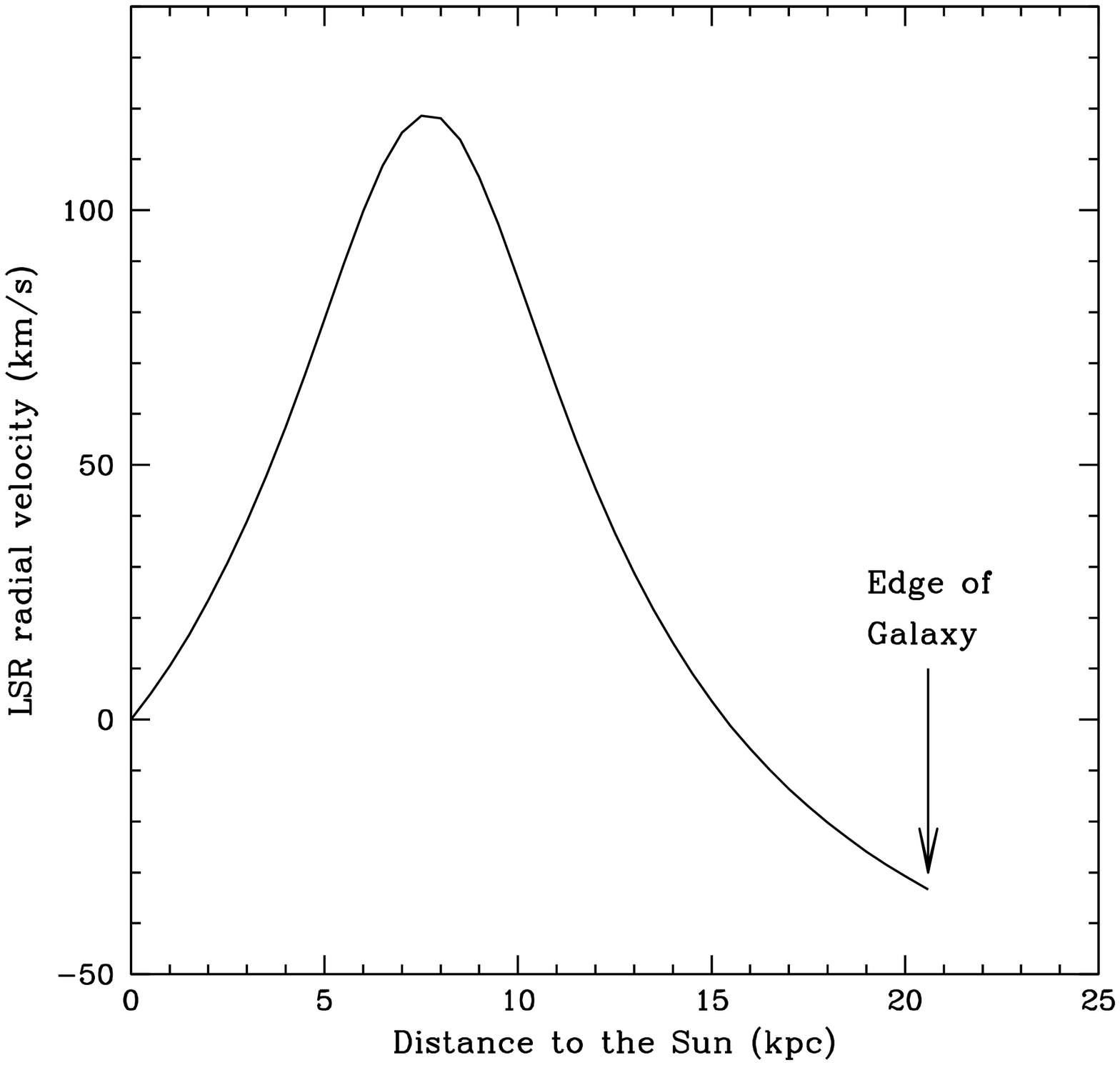}
  \caption{LSR radial velocity as a function of distance to the Sun
in the direction of the cluster of red supergiants. The negative values
of the LSR radial velocity come from gas beyond the solar circle.
The arrow marks the distance to the Sun where the edge of the Galaxy is
located.}
  \label{fig:distance}
\end{figure}       

\subsection{Distances}

Figer et al. (2006) estimate the distance to the cluster of red 
supergiants to be 5.8 kpc. For this, they associate a 1612 MHz OH maser
(OH~25.3$-$0.16)
with LSR radial velocity of +102.2 km s$^{-1}$ with the cluster.
This association is well justified, since the maser is located near the
center of the cluster. However, these masers are usually produced in the winds of
evolved stars and typically exhibit a double-peaked spectrum, with the systemic velocity
of the star at the middle of the velocity of the two lines.
In the case of OH~25.3$-$0.16
only one of the lines was detected, implying that the true systemic velocity 
of the star is either redshifted or blueshifted with respect to the
value of +102.2 km s$^{-1}$. Additional observations of this maser
are required to refine the distance to the cluster.

In the case of the H~II region G25.38-0.18, the lack of absorption at
negative velocities (see Fig. 2) indicates it is inside the solar circle.
Furthermore, the lack of HI absorption at LSR velocities more positive than
+60 km s$^{-1}$ suggests a distance of 4.1 kpc. This H~II region
has reported H76$\alpha$ emission at an LSR velocity of
+58.9$\pm$0.4 km s$^{-1}$, implying a distance of also 4.1 kpc.

In contrast, the HI absorption at negative LSR velocities of order 
$-$30 km s$^{-1}$ present in the three non-thermal sources (see Fig. 2) indicates
that they are outside the solar circle. From Figure 3 we see that the
expected LSR velocity for gas at the outer edge of the Galaxy in that
direction is also $-$30 km s$^{-1}$, implying that the sources are outside of the
Galaxy, most probably at extragalactic distances.

\section{Conclusions}

We have presented an analysis of the HI absorption spectra toward three
bright non thermal radio sources associated in the plane of the sky with
the extraordinarily massive cluster of red supergiants
recently reported by Figer et al. (2006).
Even when the \sl a priori \rm probability that these sources are
background extragalactic objects is very small, the LSR velocity range of
the HI absorption seems to indicate this is the case.
We thus confirm the classification of these sources as extragalactic, as first
noted by Kolpak et al. (2002). 
Further research is required to understand the unlikely association of this
remarkable cluster with radio, X-ray, and $\gamma$-ray sources.


\acknowledgments
LFR acknowledges the support
of DGAPA, UNAM, and of CONACyT (M\'exico).
This research has made use of the SIMBAD database, 
operated at CDS, Strasbourg, France.


\end{document}